\documentclass[10pt,conference]{IEEEtran}
\usepackage{cite}
\usepackage{amsthm}
\usepackage{graphicx}
\usepackage{stfloats}
\usepackage[cmex10]{amsmath}
\usepackage{algorithmic}
\usepackage{array}
\usepackage{mdwmath}
\usepackage{mdwtab}
\usepackage{amsfonts}
\usepackage{amssymb}
\usepackage{eqparbox}
\usepackage{multirow}

\begin{document}
\title{Space-Frequency Block Code for MIMO-OFDM Communication Systems with Reconfigurable Antennas}
\author{\IEEEauthorblockN{Vida Vakilian\IEEEauthorrefmark{1}, Jean-Fran\c{c}ois Frigon\IEEEauthorrefmark{1},
and S\'{e}bastien Roy\IEEEauthorrefmark{2}}
\IEEEauthorblockA{\IEEEauthorrefmark{1} \'{E}cole Polytechnique de
Montr\'{e}al, Dept. of Electrical Engineering, Montr\'{e}al, QC,
H3T 1J4, Canada
\\\{vida.vakilian, j-f.frigon\}@polymtl.ca}
\IEEEauthorblockA{\IEEEauthorrefmark{2} Dept. of Elec. and Comp. Engineering, Universit\'{e} Sherbrooke
Sherbrooke, QC, Canada
\\sebastien.roy13@usherbrooke.ca}}
\maketitle
\begin{abstract}
We propose a space-frequency (SF) block coding
scheme for a multiple-input multiple-output (MIMO) orthogonal
frequency-division multiplexing (OFDM) system using antennas with
reconfigurable radiation patterns. In this system, each element of
the antenna array at the transmitter side is assumed to be 
reconfigurable so that it can independently change the
physical characteristics of its radiation pattern. The proposed
block code is full rate and benefits from spatial, frequency, and
reconfigurable radiation pattern state diversity over
frequency-selective fading channels.  We provide simulation
results to demonstrate the performance of the proposed block
coding technique and make comparisons with that of the previous SF
coding schemes in MIMO-OFDM systems. The results indicate that the proposed
code achieves higher diversity and coding gain compared to other available SF codes.
\end{abstract}


\section{Introduction}
Reconfigurable antennas can be used in multiple-input
multiple-output (MIMO) communication systems to increase the
capacity and reliability of wireless links
\cite{cetiner2006mimo,piazza2008design,frigon2008dynamic,
li2009capacity,grau2008reconfigurable}. In a reconfigurable MIMO
system, the characteristics of each antenna radiation pattern can be changed
by placing switching devices such as Microelectromechanical
Systems (MEMS), varactor diodes, or field-effect transistor (FET)
within the antenna structure
\cite{weedon2001mems,caloz2005electromagnetic,won2006reconfigurable}.
As a result, a system employing reconfigurable antennas is able to
alter the propagation characteristics of the wireless channel into
a form that leads to a better signal quality at the receiver. In
fact, by using reconfigurable antennas and designing a proper
code, we can achieve an additional diversity gain that can further
improve the performance of wireless communication systems.

There are several works in the literature on designing efficient
codes for reconfigurable MIMO systems in order to take advantage
of the antenna reconfigurability. In
\cite{grau2008reconfigurable}, authors have proposed a MIMO
system equipped with reconfigurable antennas at the receiver that can 
achieve a diversity order that equals to the product of the number
of transmit antennas, the number of receive antennas and the
number of reconfigurable states of the receive antennas. They have
shown that this diversity gain is achievable only under certain
channel propagation conditions and using an appropriate coding
technique. Later on, in \cite{fazel2009space} authors extended
the concept by using reconfigurable elements at both transmitter
and receiver sides. In their work, they have
introduced a state-switching transmission scheme, called
space-time-state block coding (STS-BC), to further utilize the
available diversity in the system over flat fading wireless
channels. However, their coding scheme does not exploit the
frequency diversity offered by the multipath propagation channels
between each transmit and receive antenna pair.

To obtain frequency diversity in multipath environment, a
space-frequency (SF) block code was first proposed by authors in
\cite{agrawal1998space}, where they used the existing space-time
(ST) coding concept and constructed the code in frequency domain.
Later works
\cite{lee2000space,bolcskei2000space,lu2000space,blum2001improved,hong2002robust,lee2000space1}
also used similar strategies to develop SF codes for MIMO-OFDM
systems. However, the resulting SF codes achieved only spatial
diversity, and they were not able to obtain both spatial and
frequency diversities. To address this problem, a subcarrier
grouping method has been proposed in \cite{molisch2002space} to
further enhance the diversity gain while reducing the receiver
complexity. In \cite{su2003obtaining}, a repetition mapping
technique has been proposed that obtains full-diversity in
frequency-selective fading channels. Although their proposed
technique achieves full-diversity order, it does not guarantee
full coding rate. Subsequently, a block coding technique that
offers full-diversity and full coding rate was derived
\cite{su2005full,fazel2008quasi}.
However, the SF codes proposed in the above studies and other
similar works on the topic are not able to exploit the state
diversity available in reconfigurable multiple antenna
systems.

In this paper, we propose a coding scheme for reconfigurable
MIMO-OFDM systems that achieves multiple diversity gains,
including, space, frequency, and state.
Basically, the proposed scheme consists of a code that is sent
over transmit antennas, OFDM tones, and radiation states. In order
to obtain state diversity, we configure each transmit antenna
element to independently switch its radiation pattern to a
direction that can be selected according to different optimization
criteria, e.g., to minimize the correlation among different
radiation states. We construct our proposed code based on the
fundamental concept of rotated quasi-orthogonal space-time block
codes (QOSTBC)
\cite{tirkkonen2001optimizing,sharma2003improved,su2004signal}.
By using the rotated QOSTBC, the proposed coding structure
provides rate-one transmission (i.e., one symbol per frequency
subcarrier per radiation state) and leads to a simpler Maximum
Likelihood (ML) decoder. As the simulation results indicate, our
proposed code outperforms the existing space-frequency codes
substantially.

\begin{figure*}
  \centering
  \includegraphics[scale=0.8]{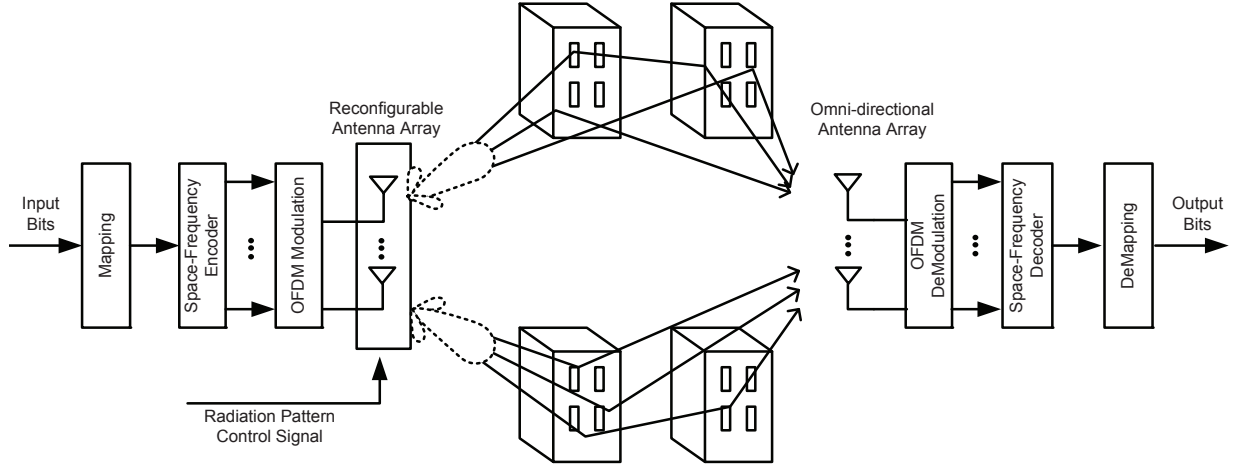}
  \caption{\footnotesize{Block diagram of a Reconfigurable MIMO-OFDM system employing reconfigurable antennas at the transmitter.}}
  \label{fig:Sys_Blk}
\end{figure*}

The rest of this paper is organized as follows. In
Section~\ref{sec:ch_sys_model}, we introduce the channel and
system model for a reconfigurable MIMO-OFDM system. In
Section~\ref{sec:code_deg}, we briefly discuss the code design for
reconfigurable multiple antenna systems. Simulation results are
presented in Section~\ref{sec:results}, and finally conclusions
are drawn in Section~\ref{sec:conc}.

\textit{Notation:} Throughout this paper, we use capital boldface
letters for matrices, and lowercase boldface letters for vectors.
$(\cdot)^T$ denotes transpose of a vector. $\mathcal{C}$ stands
for the set of complex valued numbers. Operator $\text{diag}(a_1,
a_2, \cdots, a_n)$ represents a diagonal $n \times n$ matrix whose
diagonal entries are $a_1, a_2, \cdots , a_n$. $\lfloor \cdot
\rfloor$ stands for the floor operation and ${\bf I}_N$ represents
the $N \times N$ identity matrix. Operator $\text{col}\{\cdot\}$
stacks up the matrices on top of each other.

\section{Channel and System Models for Reconfigurable MIMO-OFDM Systems}
\label{sec:ch_sys_model} Consider a MIMO-OFDM
system with $M_t$ reconfigurable elements at the transmitter
where each of these elements is capable of
electronically changing its radiation pattern and creating $P$
different radiation states as shown in Fig. \ref{fig:Sys_Blk}. In
this system, we assume the receiver antenna array consist of $M_r$
omni-directional elements with fixed radiation patterns. Moreover,
we consider an $N_c$-tone OFDM modulation and frequency-selective
fading channels with $L$ independent propagation paths between
each pair of transmit and receiver antenna in each radiation
state. The channel gains are quasi-static over one OFDM symbol
interval. The channel impulse response between transmit antenna
$i$ and receive antenna $j$ in the $p$-th radiation state can be
modelled as
\begin{align}
h_{p}^{i,j}(\tau) = \sum_{l=0}^{L-1}
\alpha_{p}^{i,j}(l)\delta(\tau-\tau_{l,p}), \label{eq:channel_t}
\end{align}
where $\tau_{l,p}$ is the $l$-th path delay in the $p$-th
radiation state, and $\alpha_{p}^{i,j}(l)$ is the complex
amplitude of the $l$-th path between the $i$-th reconfigurable
transmit antenna and the $j$-th receive antenna in the $p$-th
radiation state. The average total received power is normalized to
one.

The frequency response of the channel at the $n$-th subcarrier
between transmit antenna $i$ and receive antenna $j$ in the $p$-th
radiation state is given by
\begin{align}
H_{p}^{i,j}(n) = \sum_{l=0}^{L-1} \alpha_{p}^{i,j}(l) e^{-j2\pi n
\Delta f \tau_{l,p}},
\label{eq:channel_f}
\end{align}
where $\Delta f = 1/T_s$ is the subcarrier frequency spacing and
$T_s$ is the OFDM symbol duration. The space-frequency codeword
transmitted during the $p$-th radiation state, ${\bf C}_p \in
\mathcal{C}^{M_t \times N_c}$, can be expressed as

\begin{equation}
{\bf C}_{p} =
\left[
\begin{array}{cccc}
  c_p^1(0)     & c_p^1(1)     & \cdots &  c_p^1(N_c-1)   \\
  c_p^2(0)     & c_p^2(1)     & \cdots &  c_p^2(N_c-1)   \\
  \vdots       & \vdots       & \ddots &  \vdots         \\
  c_p^{M_t}(0) & c_p^{M_t}(1) & \cdots &  c_p^{M_t}(N_c-1)
\end{array}
\right],
\label{eq:SF}
\end{equation}
where $c_{p}^i(n)$ denotes the data symbol transmitted by transmit
antenna $i$ on the $n$-th subcarrier during the $p$-th radiation
state.

At the receiver, after cyclic prefix removal and FFT, the received
frequency domain signal of the $n$-th subcarrier and $p$-th
radiation state at the $j$-th receive antenna can be written as

\begin{eqnarray}
y_{p}^j (n) =  \sqrt{\frac{\gamma}{M_t}} \sum_{i=1}^{M_t}
H_{p}^{i,j}(n) c_{p}^i(n) + z_{p}^j(n),
\label{eq:Rx_Signal}
\end{eqnarray}
where $H_{p}^{i,j}(n)$ is the frequency response of the channel at
the $n$-th subcarrier between transmit antenna $i$ and receive
antenna $j$ in the $p$-th radiation state as defined in
(\ref{eq:channel_f}), $z_{p}^j(n)$ is the additive complex
Gaussian noise with zero mean and unit variance at the $n$-th
subcarrier, and $\gamma$ is the received signal-to noise ratio
(SNR).

The received signal during the $p$-th radiation state ${\bf y}_p =
[{\bf y}_p^T(0) \; {\bf y}_p^T(1) \; \cdots \; {\bf
y}_p^T(N_c-1)]^T$ with ${\bf y}_p(n) = [y_{p}^1 (n) \; y_{p}^2 (n)
\; \cdots \; y_{p}^{M_r} (n)]^T$, can be written as

\begin{eqnarray}
{\bf y}_p = \sqrt{\frac{\gamma}{M_t}} {\bf H}_p {\bf c}_p + {\bf z}_p,
\label{eq:Rx_Signal_state}
\end{eqnarray}
where
\begin{equation}
{\bf H}_p = \text{diag}\{{\bf H}_p(0),\; {\bf H}_p(1),\; \cdots,\;
{\bf H}_p(N_c-1)\} \label{eq:channel_matrix_state}
\end{equation}
is the channel matrix, ${\bf c}_p = \text{vec} ({\bf C}_p)$ is the
transmitted codeword, and ${\bf z}_p \in \mathcal{C}^{N_cM_r
\times 1}$ is the noise vector during the $p$-th radiation state.
In (\ref{eq:channel_matrix_state}), ${\bf H}_p(n)$ is an $M_r
\times M_t$ channel matrix with entries defined in
(\ref{eq:channel_f}).

The SF codeword over all $P$ radiation states can be represented
as
\begin{equation}
{\bf C} = \big[{\bf C}_1, \; {\bf C}_2,\;\cdots, \; {\bf
C}_P\big], \label{eq:Code}
\end{equation}
where ${\bf C}_p$ is given in (\ref{eq:SF}). The received signals
over all radiation states is defined by ${\bf y} = [{\bf y}_1^T \;
{\bf y}_2^T \; \cdots \; {\bf y}_P^T]^T \in \mathcal{C}^{PN_c M_r
\times 1}$ and can be represented by

\begin{eqnarray}
{\bf y} = \sqrt{\frac{\gamma}{M_t}} {\bf H} {\bf c} + {\bf z},
\label{eq:Rx_Signal_state}
\end{eqnarray}
where ${\bf c} = \text{vec} ({\bf C})$, ${\bf H} = \text{diag}
\{{\bf H}_1,\; {\bf H}_2,\; \cdots,\; {\bf H}_P\} \in
\mathcal{C}^{PN_cM_r \times PN_cM_t} $ is the overall channel
matrix, and ${\bf z} = [{\bf z}_1^T \; {\bf z}_2^T \; \cdots \;
{\bf z}_P^T]^T \in \mathcal{C}^{PN_c M_r \times 1}$ is the noise
vector.

\newcounter{tempequationcounter}
\begin{figure*}[b!]
\setcounter{tempequationcounter}{\value{equation}} \hrulefill
\vspace*{4pt}
\begin{equation*}
\setcounter{equation}{13} {\bf C}_1 = \frac{1}{4} \left[
  \begin{array}{ccccccccc}
   \mathcal{S}_1^1  & -\mathcal{S}^{1^*}_2 & \mathcal{S}^1_3 & -\mathcal{S}^{1^*}_4 & \cdots & \mathcal{S}_1^M & -\mathcal{S}^{M^*}_2 & \mathcal{S}^M_3  & -\mathcal{S}^{M^*}_4 \\
   \mathcal{S}_2^1  &  \mathcal{S}^{1*}_1  & \mathcal{S}^1_4 &  \mathcal{S}^{1*}_3  & \cdots & \mathcal{S}_2^M & \mathcal{S}^{M*}_1   & \mathcal{S}^M_4  & \mathcal{S}^{M*}_3   \\
  \end{array}
\right]
\end{equation*}
\begin{equation}
{\bf C}_2 = \frac{1}{4} \left[
  \begin{array}{ccccccccc}
   \mathcal{S}_5^1  & -\mathcal{S}^{1^*}_6 & \mathcal{S}^1_7 & -\mathcal{S}^{1^*}_8 & \cdots & \mathcal{S}_5^M & -\mathcal{S}^{M^*}_6 & \mathcal{S}^M_7  & -\mathcal{S}^{M^*}_8 \\
   \mathcal{S}_6^1  &  \mathcal{S}^{1*}_5  & \mathcal{S}^1_8 &  \mathcal{S}^{1*}_7  & \cdots & \mathcal{S}_6^M & \mathcal{S}^{M*}_5   & \mathcal{S}^M_8  & \mathcal{S}^{M*}_7   \\
  \end{array}
\right] \label{eq:codewordP4}
\end{equation}
\setcounter{equation}{\value{tempequationcounter}}
\end{figure*}

\section{Space-Frequency Code Design for Reconfigurable MIMO-OFDM Systems}
\label{sec:code_deg} In this section, we present our proposed
coding scheme for a reconfigurable antenna system, where each
antenna elements can independently change its radiation pattern
direction. In particular, we construct the code based on the
principle of a quasi-orthogonal coding structure for an arbitrary
number of transmit antennas and radiation pattern states. In each
radiation state, we consider a coding strategy where the SF
codeword is a concatenation of ${\bf G}_p^{m^T}$ as follows:

\begin{equation}
{\bf C}_p = [ {\bf G}_p^{1^T} {\bf G}_p^{2^T} \cdots  {\bf
G}_p^{M^T} {\bf 0}^T_{N_c-MLM_t} ], \label{eq:SFS_code}
\end{equation}
where $M=\lfloor \frac{N_c}{LM_t}\rfloor$ and ${\bf 0}_N$ is the
all-zeros $N \times N$ matrix. In this expression, ${\bf 0}_N$
will disappear if $N_c$ is an integer multiple of $LM_t$. In this
work, for simplicity, we assume $N_c=LM_tq$, for some integer $q$.
Each ${\bf G}_p^{m}$ matrix, $m \in \{1,2,\cdots,M\}$, takes the
following form:

\begin{equation}
{\bf G}_p^{m} = \text{col}\{ {\bf X}_1, \; {\bf X}_2, \; \cdots,
\; {\bf X}_L\},
\end{equation}
where ${\bf X}_l$ is the $M_t \times M_t$ block coding matrix
which is equivalent to an Alamouti code structure for $M_t=2$. To
maintain simplicity in our presentation, we design the code for
$M_t=2$ transmit antennas, however, extension to $M_t>2$ is
possible by following the similar procedure. In the case of having
two transmit antennas, ${\bf X}_l = {\bf A}\big(x_1, x_2\big)$,
where

\begin{equation}
{\bf A}\big(x_1, x_2\big) = \left[ {\begin{array}{cc} x_1 & x_2 \\
-x_2^* & x_1^* \\ \end{array} } \right],
\end{equation}
and therefore ${\bf G}_p^{m}$ can be expressed as

\begin{equation}
{\bf G}_p^{m} = \left[
\begin{array}{c}
{\bf A}(\mathcal{S}_{2(p-1)L+1}^m,\mathcal{S}_{2(p-1)L+2}^m)\\
{\bf A}(\mathcal{S}_{2(p-1)L+3}^m,\mathcal{S}_{2(p-1)L+4}^m)\\
\vdots \\
{\bf A}(\mathcal{S}_{2pL-1}^m,\mathcal{S}_{2pL}^m)\\
\end{array}
\right].
\label{eq:G}
\end{equation}
In (\ref{eq:G}), $\mathcal{S}_i^m$ is a set of combined symbols,
computed as
\begin{align}
&\left[
\begin{array}{cccc}
\mathcal{S}_{\scriptscriptstyle 1}^m
\;\mathcal{S}_{\scriptscriptstyle 3}^m \;\cdots
\;\mathcal{S}_{\scriptscriptstyle 2PL-1}^m
\end{array}
\right]^T = \Theta \left[
\begin{array}{cccc}
s_{\scriptscriptstyle 1}^m \;s_{\scriptscriptstyle 3}^m \;\cdots
\; s_{\scriptscriptstyle 2PL-1}^m
\end{array}
\right]^T, \nonumber\\
&\left[
\begin{array}{cccc}
\mathcal{S}_{\scriptscriptstyle 2}^m
\;\mathcal{S}_{\scriptscriptstyle 4}^m \;\cdots
\;\mathcal{S}_{\scriptscriptstyle 2PL}^m\;\;\;
\end{array}
\right]^T = \Theta \left[
\begin{array}{cccc}
s_{\scriptscriptstyle 2}^m \;s_{\scriptscriptstyle 4}^m \;\cdots
\;s_{\scriptscriptstyle 2PL}^m\;\;\;
\end{array}
\right]^T, \label{eq:codewordP3}
\end{align}
where $\{s_{\scriptscriptstyle 1}^m,\cdots,s_{\scriptscriptstyle
2PL}^m\}$ is a block of symbols belonging to a constellation
$\mathcal{A}$, $\Theta = {\bf U} \times
\text{diag}\{1,e^{j\theta_1},\ldots,e^{j\theta_{PL-1}}\}$ and
${\bf U}$ is a $PL \times PL$ Hadamard matrix. The $\theta_i$'s
are the rotation angles. Different optimization strategies can be
used to find the optimal values of rotation angles $\theta_i$'s,
such that they maximize the coding gain. The objective function in
this optimization is defined as the minimum Euclidean distance
between constellation points.

As an example, consider a reconfigurable MIMO-OFDM system with
$M_t=2$ transmit antennas, $P=2$ radiation states, and $L=2$
multipaths. In this scenario, the transmitted codewords ${\bf
C}_1$ and ${\bf C}_2$ given in (\ref{eq:codewordP4}) are
constructed according to (\ref{eq:SFS_code}). The entries of ${\bf
C}_p$ are computed using (\ref{eq:codewordP3}). As a result, we
obtain ${\bf C}_1^T$ as

\setcounter{equation}{14} \small
\begin{align}
\left[
\begin{array}{cc}
s_1^1+\tilde{s}_3^1+\hat{s}_5^1+\check{s}_7^1                  & s_2^1+\tilde{s}_4^1+\hat{s}_6^1+\check{s}_8^1                  \\
-s_2^{1^*}-\tilde{s}_4^{1^*}-\hat{s}_6^{1^*}-\check{s}_8^{1^*} & s_1^{1^*}+\tilde{s}_3^{1^*}+\hat{s}_5^{1^*}+\check{s}_7^{1^*}  \\
s_1^1-\tilde{s}_3^1+\hat{s}_5^1-\check{s}_7^1                  & s_2^1-\tilde{s}_4^1+\hat{s}_6^1-\check{s}_8^1                  \\
-s_2^{1^*}+\tilde{s}_4^{1^*}-\hat{s}_6^{1^*}+\check{s}_8^{1^*} & s_1^{1^*}-\tilde{s}_3^{1^*}+\hat{s}_5^{1^*}-\check{s}_7^{1^*}  \\
\vdots & \vdots\\
s_1^M+\tilde{s}_3^M+\hat{s}_5^M+\check{s}_7^M                  & s_2^M+\tilde{s}_4^M+\hat{s}_6^M+\check{s}_8^M                  \\
-s_2^{M^*}-\tilde{s}_4^{M^*}-\hat{s}_6^{M^*}-\check{s}_8^{M^*} & s_1^{M^*}+\tilde{s}_3^{M^*}+\hat{s}_5^{M^*}+\check{s}_7^{M^*}  \\
s_1^M-\tilde{s}_3^M+\hat{s}_5^M-\check{s}_7^M                  & s_2^M-\tilde{s}_4^M+\hat{s}_6^M-\check{s}_8^M                  \\
-s_2^{M^*}+\tilde{s}_4^{M^*}-\hat{s}_6^{M^*}+\check{s}_8^{M^*} & s_1^{M^*}-\tilde{s}_3^{M^*}+\hat{s}_5^{M^*}-\check{s}_7^{M^*}  \\
\end{array}
\right], \label{eq:codewordP5}
\end{align}
and ${\bf C}_2^T$ as \small
\begin{equation}
\left[
\begin{array}{cc}
s_1^1+\tilde{s}_3^1-\hat{s}_5^1-\check{s}_7^1                  & s_2^1+\tilde{s}_4^1-\hat{s}_6^1-\check{s}_8^1                  \\
-s_2^{1^*}-\tilde{s}_4^{1^*}+\hat{s}_6^{1^*}+\check{s}_8^{1^*} & s_1^{1^*}+\tilde{s}_3^{1^*}-\hat{s}_5^{1^*}-\check{s}_7^{1^*}  \\
s_1^1-\tilde{s}_3^1-\hat{s}_5^1+\check{s}_7^1                  & s_2^1-\tilde{s}_4^1-\hat{s}_6^1+\check{s}_8^1                  \\
-s_2^{1^*}+\tilde{s}_4^{1^*}+\hat{s}_6^{1^*}-\check{s}_8^{1^*} & s_1^{1^*}-\tilde{s}_3^{1^*}-\hat{s}_5^{1^*}+\check{s}_7^{1^*}  \\
\vdots & \vdots\\
s_1^M+\tilde{s}_3^M-\hat{s}_5^M-\check{s}_7^M                  & s_2^M+\tilde{s}_4^M-\hat{s}_6^M-\check{s}_8^M                  \\
-s_2^{M^*}-\tilde{s}_4^{M^*}+\hat{s}_6^{M^*}+\check{s}_8^{M^*} & s_1^{M^*}+\tilde{s}_3^{M^*}-\hat{s}_5^{M^*}-\check{s}_7^{M^*}  \\
s_1^M-\tilde{s}_3^M-\hat{s}_5^M+\check{s}_7^M                  & s_2^M-\tilde{s}_4^M-\hat{s}_6^M+\check{s}_8^M                  \\
-s_2^{M^*}+\tilde{s}_4^{M^*}+\hat{s}_6^{M^*}-\check{s}_8^{M^*} & s_1^{M^*}-\tilde{s}_3^{M^*}-\hat{s}_5^{M^*}+\check{s}_7^{M^*}  \\
\end{array}
\right], \label{eq:codewordP6}
\end{equation}
\normalsize
where $\tilde{s}_i = e^{j\theta_1}s_i$, $\hat{s}_i =
e^{j\theta_2}s_i$, and $\check{s}_i = e^{j\theta_3}s_i$. Note that
the above codeword provides rate-one transmission (i.e., one
symbol per OFDM tone per radiation state).

\section{Simulation Results}
\label{sec:results} In this section, we present simulation results
for both conventional and reconfigurable MIMO-OFDM systems. The
reconfigurable multiple antenna system employs antenna elements
capable of dynamically changing their radiation pattern directions
at the transmitter, where in this work, we consider that each
element has $P = 2$ radiation states. The conventional MIMO-OFDM
system uses omni-directional antenna elements with fixed radiation
pattern at both transmitter and receiver ends. For both systems,
we consider $M_t = 2$ antennas at the transmitter and $M_r = 1$
antenna at the receiver and an OFDM modulation technique with $N_c
= 128$ subcarriers as well as a cyclic prefix equal to or longer
than the maximum channel delay spread. In our simulations, we
consider that the receiver has perfect channel state information.
We also assume that the symbols are chosen from a BPSK
constellation, leading to a spectral efficiency of 1 bit/sec/Hz if
the cyclic prefix overhead is ignored. The average symbol power
per transmit antenna is set to be $E_s = 1/M_t$ and the noise
variance is $\sigma_n^2 = 1/\gamma$. We carry out the simulations
for a $2$-ray equal power channel model for two different delay
spreads. Furthermore, for a reconfigurable antenna system, we
assume the same delay spread for both radiation states (i.e.,
$\tau_{l,1} = \tau_{l,2}$). The channel coefficients
$\alpha_{p}^{i,j}(l)$ are zero-mean identically-distributed
Gaussian random variables with a variance of $\sigma_{l,p}^2$. We
assumed that they are independent for each multipath, transmit
antenna and radiation pattern state. The powers of all paths in
each radiation state are normalized such that $\sum_{l=0}^{L-1}
\sigma_{l,p}^2 = 1$. For our proposed QOSF scheme, the rotation
angles are chosen as $\theta_1 = \frac{\pi}{4}$, $\theta_2 =
\frac{\pi}{2}$, and $\theta_3 = \frac{3\pi}{4}$.

\begin{figure}
\centering
\includegraphics [width=3.8in]{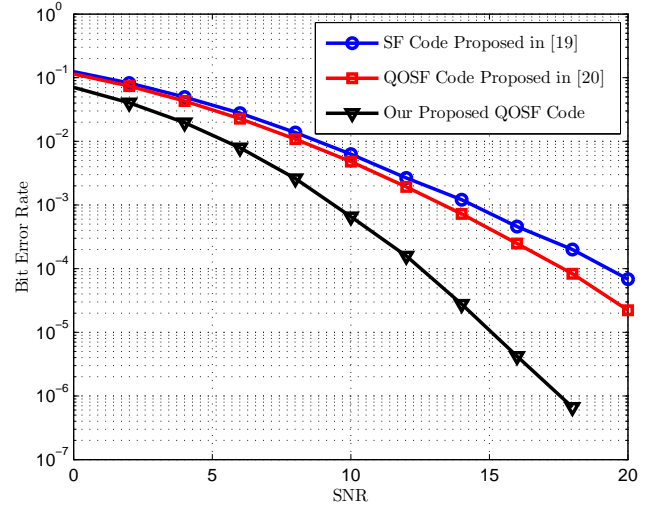}
\caption{BER vs. SNR for a reconfigurable multi-antenna system
with $M_t=2$, $P = 2$, $M_r=1$ in a $2$-ray channel with a delay
spread of $5 \mu s$} \label{fig:BER_tau5}
\end{figure}

Fig. \ref{fig:BER_tau5} shows the bit error rate (BER) performance
of the proposed code in multipath propagation channels with a
delay spread of $\tau = 5 \mu s$.
As shown in this figure, the proposed code outperforms those of
\cite{su2005full} and \cite{fazel2008quasi}. In particular, at a
bit error rate of $10^{-5}$, the performance improvement compared
to \cite{su2005full} and \cite{fazel2008quasi} is nearly $7$ and
$6$ dB, respectively. This performance improvement demonstrates
the superiority of our proposed scheme which is due to the extra
diversity gain offered by the use of reconfigurable antenna
elements.

\begin{figure}
\centering
\includegraphics [width=3.8in]{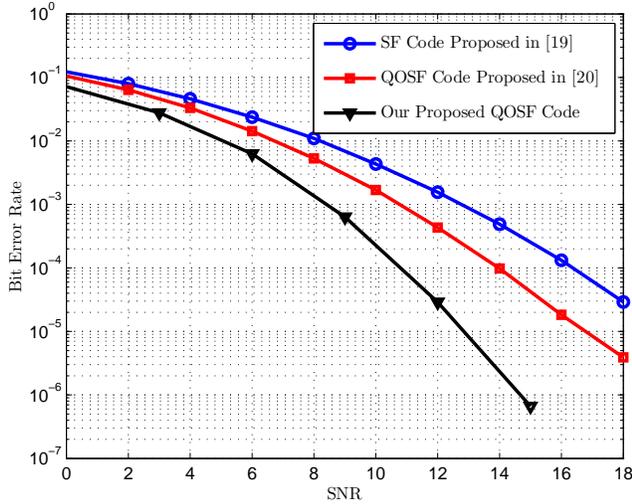}
\caption{BER vs. SNR for a reconfigurable multi-antenna system
with $M_t=2$, $P = 2$, $M_r=1$ in a $2$-ray channel with a delay
spread of $20 \mu s$ } \label{fig:BER_tau20}
\end{figure}

Fig. \ref{fig:BER_tau20} depicts the BER performance of the
proposed code for a delay spread of $\tau = 20 \mu s$. It is
evident from the figure that at a BER of $10^{-5}$, our proposed
coding scheme outperforms the codes presented in \cite{su2005full}
and \cite{fazel2008quasi} by about $6$ and $4$ dB, respectively.
Compared to the results in Fig. \ref{fig:BER_tau5}, it can be seen
that as delay spread increases, the BER performance improves. This
is due to benefiting from lower correlation between subcarriers,
and therefore higher frequency diversity in multipath propagation
channels.

\section{Conclusions}
\label{sec:conc} We proposed a space-frequency coding technique
for MIMO-OFDM systems using antennas with reconfigurable radiation
patterns. The proposed code is constructed based on the principle
of quasi-orthogonal coding scheme and consists of a block of
transmitted symbols expanding over space, frequency, and radiation
state dimensions. We provided simulation results to demonstrate
the performance of the proposed coding scheme and make comparisons
with that of the previous SF coding schemes. In these experiments,
it has been shown that the proposed code provides additional
diversity and coding gains compared to the previously designed SF
codes in MIMO-OFDM systems.

\bibliographystyle{IEEEtran}
\bibliography{IEEEabrv,Reference}

\begin{thebibliography}{10}
\providecommand{\url}[1]{#1}
\csname url@rmstyle\endcsname
\providecommand{\newblock}{\relax}
\providecommand{\bibinfo}[2]{#2}
\providecommand\BIBentrySTDinterwordspacing{\spaceskip=0pt\relax}
\providecommand\BIBentryALTinterwordstretchfactor{4}
\providecommand\BIBentryALTinterwordspacing{\spaceskip=\fontdimen2\font plus
\BIBentryALTinterwordstretchfactor\fontdimen3\font minus
  \fontdimen4\font\relax}
\providecommand\BIBforeignlanguage[2]{{%
\expandafter\ifx\csname l@#1\endcsname\relax
\typeout{** WARNING: IEEEtran.bst: No hyphenation pattern has been}%
\typeout{** loaded for the language `#1'. Using the pattern for}%
\typeout{** the default language instead.}%
\else
\language=\csname l@#1\endcsname
\fi
#2}}

\bibitem{cetiner2006mimo}
B.~Cetiner, E.~Akay, E.~Sengul, and E.~Ayanoglu, ``A {MIMO} system with
  multifunctional reconfigurable antennas,'' \emph{IEEE Antennas Wireless
  Propag. Lett.}, vol.~5, pp. 463--466, 2006.

\bibitem{piazza2008design}
D.~Piazza, N.~Kirsch, A.~Forenza, R.~Heath, and K.~Dandekar, ``{Design and
  evaluation of a reconfigurable antenna array for {MIMO} systems},''
  \emph{{IEEE} Trans. Antennas Propagat.}, vol.~56, pp. 869--881, Mar. 2008.

\bibitem{frigon2008dynamic}
J.~Frigon, C.~Caloz, and Y.~Zhao, ``{Dynamic radiation pattern diversity
  {(DRPD) MIMO} using {CRLH} leaky-wave antennas},'' in \emph{Proc. IEEE Radio
  and Wireless Symp.}, 2008, pp. 635--638.

\bibitem{li2009capacity}
X.~Li and J.~Frigon, ``Capacity analysis of {MIMO} systems with dynamic
  radiation pattern diversity,'' in \emph{Proc. IEEE VTC Spring 2009}, 2009,
  pp. 1--5.

\bibitem{grau2008reconfigurable}
A.~Grau, H.~Jafarkhani, and F.~De~Flaviis, ``A reconfigurable multiple-input
  multiple-output communication system,'' \emph{{IEEE} Trans. on Wireless
  Commun.}, vol.~7, pp. 1719--1733, May 2008.

\bibitem{weedon2001mems}
W.~Weedon, W.~Payne, and G.~Rebeiz, ``{MEMS}-switched reconfigurable
  antennas,'' in \emph{Antennas and Propagation Society International
  Symposium, 2001. IEEE}, vol.~3, 2001, pp. 654--657.

\bibitem{caloz2005electromagnetic}
C.~Caloz and T.~Itoh, \emph{Electromagnetic metamaterials: transmission line
  theory and microwave applications}.\hskip 1em plus 0.5em minus 0.4em\relax
  Wiley-IEEE Press, 2005.

\bibitem{won2006reconfigurable}
C.~won Jung, M.-j. Lee, G.~Li, and F.~De~Flaviis, ``Reconfigurable scan-beam
  single-arm spiral antenna integrated with {RF-MEMS} switches,'' \emph{{IEEE}
  Trans. Antennas Propagat.}, vol.~54, pp. 455--463, Feb. 2006.

\bibitem{fazel2009space}
F.~Fazel, A.~Grau, H.~Jafarkhani, and F.~Flaviis, ``Space-time-state block
  coded {MIMO} communication systems using reconfigurable antennas,''
  \emph{{IEEE} Trans. on Wireless Commun.}, vol.~8, pp. 6019--6029, Dec. 2009.

\bibitem{agrawal1998space}
D.~Agrawal, V.~Tarokh, A.~Naguib, and N.~Seshadri, ``Space-time coded {OFDM}
  for high data-rate wireless communication over wideband channels,'' in
  \emph{Proc. IEEE Veh. Technol. Conf. ({VTC})}, vol.~3, 1998, pp. 2232--2236.

\bibitem{lee2000space}
K.~F. Lee and D.~B. Williams, ``A space-time coded transmitter diversity
  technique for frequency selective fading channels,'' in \emph{in Proc. IEEE
  Sensor Array and Multichannel Signal Processing Workshop}, March 2000, pp.
  149--152.

\bibitem{bolcskei2000space}
H.~Bolcskei and A.~J. Paulraj, ``Space-frequency coded broadband {OFDM}
  systems,'' in \emph{Proc. IEEE Wireless Commun. Networking Conf. ({WCNC})},
  vol.~1, Sept. 2000, pp. 1--6.

\bibitem{lu2000space}
B.~Lu and X.~Wang, ``Space-time code design in {OFDM} systems,'' in \emph{Proc.
  IEEE Global Commun. Conf. ({GLOBECOM})}, vol.~2, Nov. 2000, pp. 1000--1004.

\bibitem{blum2001improved}
R.~S. Blum, Y.~G. Li, J.~H. Winters, and Q.~Yan, ``Improved space-time coding
  for {MIMO-OFDM} wireless communications,'' \emph{{IEEE} Trans. Commun.},
  vol.~49, pp. 1873--1878, Nov. 2001.

\bibitem{hong2002robust}
Z.~Hong and B.~L. Hughes, ``Robust space-time codes for broadband {OFDM}
  systems,'' in \emph{Proc. IEEE Wireless Commun. Networking Conf. ({WCNC})},
  vol.~1, 2002, pp. 105--108.

\bibitem{lee2000space1}
K.~F. Lee and D.~B. Williams, ``A space-frequency transmitter diversity
  technique for {OFDM} systems,'' in \emph{Proc. IEEE Global Commun. Conf.
  ({GLOBECOM})}, vol.~3, 2000, pp. 1473--1477.

\bibitem{molisch2002space}
A.~F. Molisch, M.~Z. Win, and J.~H. Winters, ``Space-time-frequency ({STF})
  coding for {MIMO-OFDM} systems,'' \emph{{IEEE} Commun. Lett.}, vol.~6, pp.
  370--372, Sept. 2002.

\bibitem{su2003obtaining}
W.~Su, Z.~Safar, M.~Olfat, and K.~R. Liu, ``Obtaining full-diversity
  space-frequency codes from space-time codes via mapping,'' \emph{{IEEE}
  Trans. Signal Process.}, vol.~51, no.~11, pp. 2905--2916, Nov. 2003.

\bibitem{su2005full}
W.~Su, Z.~Safar, and K.~R. Liu, ``Full-rate full-diversity space-frequency
  codes with optimum coding advantage,'' \emph{{IEEE} Trans. Inf. Theory},
  vol.~51, pp. 229--249, Jan. 2005.

\bibitem{fazel2008quasi}
F.~Fazel and H.~Jafarkhani, ``Quasi-orthogonal space-frequency and
  space-time-frequency block codes for {MIMO} {OFDM} channels,'' \emph{{IEEE}
  Trans. on Wireless Commun.}, vol.~7, pp. 184--192, Jan. 2008.

\bibitem{tirkkonen2001optimizing}
O.~Tirkkonen, ``Optimizing space-time block codes by constellation rotations,''
  in \emph{Finnish Wireless Commun. Workshop (FWWC)}, Oct. 2001, pp. 59--60.

\bibitem{sharma2003improved}
N.~Sharma and C.~B. Papadias, ``Improved quasi-orthogonal codes through
  constellation rotation,'' \emph{{IEEE} Trans. Commun.}, vol.~51, pp.
  332--335, March 2003.

\bibitem{su2004signal}
W.~Su and X.-G. Xia, ``Signal constellations for quasi-orthogonal space-time
  block codes with full diversity,'' \emph{{IEEE} Trans. Inf. Theory}, vol.~50,
  pp. 2331--2347, Oct. 2004.

\end{thebibliography}

\end{document}